\documentclass[aps,prl,twocolumn,superscriptaddress,noeprint,longbibliography, nofootinbib,floatfix]{revtex4-2}

\usepackage{graphicx}
\usepackage{bm}
\usepackage{amsmath, amssymb}
\usepackage{booktabs}
\usepackage{amsfonts}
\usepackage{siunitx}
\usepackage{mathrsfs}
\usepackage[normalem]{ulem}
\usepackage{pifont}
\usepackage[mathscr]{euscript}
\usepackage{hyperref}
\usepackage{soul}
\usepackage{dsfont}
\usepackage[inline]{enumitem} %
\usepackage[dvipsnames]{xcolor}
\usepackage[sectionbib]{bibunits}

\definecolor{DarkerGray}{HTML}{a1a1a1}
\definecolor{LiteGray}{HTML}{e7e7e7}

\hypersetup{
    colorlinks = true,
    linkcolor  = blue,
    citecolor  = blue,
    urlcolor   = blue,
    linktocpage=true
}

\usepackage{physics}
\usepackage{comment}

\begin{document}

\title{Solving the fractional quantum Hall problem with self-attention neural network}

\author{Yi Teng}
\email{yiteng0116@gmail.com}
\affiliation{TCM Group, Cavendish Laboratory, University of Cambridge, Cambridge CB3 0HE, UK}
\affiliation{Max Planck Institute for the Physics of Complex Systems, Nöthnitzer Straße 38, 01187 Dresden, Germany}

\author{David D. Dai}
\affiliation{Department of Physics, Massachusetts Institute of Technology, Cambridge, Massachusetts 02139, USA}

\author{Liang Fu}
\email{liangfu@mit.edu}
\affiliation{Department of Physics, Massachusetts Institute of Technology, Cambridge, Massachusetts 02139, USA}

\begin{abstract}
We introduce an attention-based fermionic neural network (FNN) to variationally solve the problem of two-dimensional  Coulomb electron gas in magnetic fields, a canonical platform for fractional quantum Hall (FQH) liquids, Wigner crystals and other unconventional electron states. Working directly with the full Hilbert space of $N$ electrons confined to a disk, our FNN consistently attains energies lower than LL-projected exact diagonalization (ED) and learns the ground state wavefunction to high accuracy. In low LL mixing regime, our FNN reveals microscopic features in the short-distance behavior of FQH wavefunction beyond the Laughlin ansatz. For moderate and strong LL mixing parameters, the FNN outperforms ED significantly. Moreover, a phase transition from FQH liquid to a crystal state is found at strong LL mixing. Our study demonstrates unprecedented power and universality of FNN based variational method for solving strong-coupling many-body problems with  topological order and electron fractionalization.
\end{abstract}

\maketitle
%
%

 \section{I. Introduction}
 Theoretical quantum condensed matter physics has been driven by the desire to understand and predict complex collective phenomena---such as superconductivity and magnetism---that emerge from interactions between a large number of constituent particles. Due to the exponential growth of Hilbert space dimension with the system size, it is in general not possible to obtain numerically exact solution in the thermodynamic limit. A variety of approximate numerical methods have been developed to study different types of quantum many-body systems. The density-functional theory is remarkably successful for studying real materials with weak electron correlation effects~\cite{Kohn_1996, Perdew_2003, Becke_2014}. For strongly correlated electron systems, variational methods based on matrix product states have proven accurate for (quasi) one-dimensional lattice models~\cite{Ostlund_1995, Stoudenmire_2012, Karrasch_2012}. 

In recent years, a wide array of correlated and topological electron states have been discovered in moir\'e materials, including spin-polaron metals~\cite{Davydova_2023, Zhang_2023, Tao_2023} and fractional Chern insulators~\cite{Li3_2021, Crpel_2023, Cai_2023, Zeng_2023, Park_2023, Lu_2024}. As these ground states cannot be adiabatically connected to any mean-field states, theoretical progress has largely come from exact diagonalization (ED) studies, which are limited by small system sizes and truncation to the lowest one or few bands~\cite{Reddy_2023, Abouelkomsan_2024, Yu_2024}. To advance precision many-body theory, it is highly desirable to develop more powerful and versatile methods for solving strongly interacting electron problems in {\it continuum} systems.

Recent advances in artificial intelligence and deep learning have motivated the development of neural network-variational Monte Carlo (NN-VMC) method~\cite{Carleo_2019, Hermann_2023}. While NN-VMC approach was first introduced and developed for lattice models~\cite{Carrasquilla_2017, Luo_2019, Carleo_2019_NetKet, Kaubruegger_2018, Glasser_2018, Roth_2023}, an important breakthrough came in continuum systems when fermionic neural networks (FNN) based on generalized Slater determinants, in particular PauliNet~\cite{Hermann_2020} and FermiNet~\cite{Pfau_2020, Spencer_2020}, were developed to solve quantum chemistry problems in atoms and molecules with high accuracy. 
The NN-VMC approach has since been adapted to extended electron systems in continuous space, including real solids~\cite{Li_2022}, uniform electron gas~\cite{Cassella_2023, Pescia_2023, Luo_2023} and moir\'e materials~\cite{Li_2024, Luo_2024}. 

As a sign-problem-free variational method, NN-VMC provides a variational upper bound on the ground state energy, and its computational cost scales polynomially with the system size~\cite{VonGlehn_2023, Li2_2023}. 
Thanks to the remarkable  expressive power of neural network wavefunctions, this method can in principle solve different phases using a unified and unbiased ansatz. So far, NN-VMC has proven successful in studying Fermi liquid~\cite{Cassella_2023, Pescia_2023} and Wigner crystals~\cite{Cassella_2023, Luo_2023}. On the other hand, it remains an open question whether the {\it standard} NN architecture (such as FermiNet)
is capable of simulating exotic quantum states beyond mean-field description, such as fractionalized quantum phases of matter. 

In this work, we adapt the self-attention FNN Psiformer~\cite{VonGlehn_2023} to study two-dimensional electron gas in  strong magnetic fields---a canonical setting and enduring source for strongly correlated electron phases such as fractional quantum Hall states and composite Fermi liquids. 
Using NN-VMC, we are able to solve this many-electron problem directly {\it in real space}, without using Landau level (LL) projection and truncation. Our FNN finds the $\nu=1/3$ fractional quantum Hall state (FQH) over a range of LL mixing parameters and consistently attains energies lower than LL-projected ED. The final FNN wavefunction exhibits phase structure and pair distribution at short distance that are clearly distinct from the Laughlin ansatz. For realistic parameters, our FNN finds substantial LL mixing in the wavefunction and significantly outperforms LL-projected ED. At very strong LL mixing parameter, we discover a phase transition from the FQH liquid to another state with distinct charge density profile and quantum number.  

Our work demonstrates the power and versatility of FNN architecture for studying fractionalized quantum matter in continuous space and solving real-world solid-state problems. Our study unlocks the immense potential of NN-VMC in tackling strong correlation and topological order with a unifying and unbiased ansatz, opening the door to new applications of AI in quantum condensed matter physics.  

%
%
%
\section{II. Laughlin wavefunction and FQH ground state} 

The standard approach to the FQH problem assumes that electron-electron interaction is weak compared to the cyclotron gap, so that LL mixing is neglected to the zeroth order approximation.   
In his groundbreaking work~\cite{Laughlin_1983}, Laughlin proposed an elegant many-body wavefunction within the lowest Landau level (LLL) subspace 
at fractional filling $\nu = 1/m$ ($m$ odd):
\begin{equation}
\Psi_L = \prod_{\{ i, j \}}(z_i - z_j)^m e^{-{\Sigma_i \lvert z_i \lvert^2}/{4\ell_B^2}}\,,
\label{eq:laughlin}
\end{equation}
where $z = x + iy$ is the standard complex coordinate, and $\ell_B$ is the magentic length.

While the Laughlin wavefunction captures the essential physics of FQH liquids~\cite{Laughlin_1983, Morf_1986}, it is {\it not} the exact ground state of 2D Coulomb electron gas under magnetic field. The key difference lies in two aspects. First, in the Laughlin wavefunction the relative angular momentum $l$ of two electrons only admits $l \geq m$ components, and the zeros are all $m^{\rm th}$ order. Under realistic interaction including Coulomb, there will be non-zero components for $l < m$, with $l = 1$ component the most dominant at short distance $<\ell_B$. As a result, each $m^{\rm th}$ order zero of the wavefunction splits into $m$ first-order zeros~\cite{Tavernier_2004, Tavernier_2006, Anisimovas_2008,Tolo_2008,Wen_2012}. 

Moreover, for many semiconductor materials under realistic magnetic fields~\cite{Sodemann_2013, Shi_2020}, the interaction energy scale is comparable to or (considerably) larger than the cyclotron gap. In this case, LL mixing becomes significant, and taking into account higher LL levels is of quantitative or even qualitative importance. 

Historically, theoretical advances on the FQH problem have benefited greatly from various numerical methods including ED~\cite{Yoshioka_1983, Halperin_1984, Haldane_1985, tsiper_2001, Wan_2002, Wan_2003}, density matrix renormalization group~\cite{Zaletel_2015} and fixed-phase diffusion Monte Carlo (DMC)~\cite{Ortiz_1993, Guclu_2005, Zhao_2018, Zhao_2023}. Despite their great success, these methods have inherent limitations. ED and DMRG are limited to relatively small system size and quasi-one-dimensional geometry respectively. Moreover, both methods are limited by Landau level truncation and therefore become ineffective at strong Landau level mixing.

Similar to our NN-VMC, the fixed-phase DMC aims to solve the problem of electron gas in magnetic fields directly in real space without Landau level projection. However, it requires an input trial wavefunction such as the Laughlin ansatz and retains its phase structure while optimizing the probability distribution. As such, the accuracy of fixed-phase DMC is uncontrolled and inherently limited by the quality of the underlying trial wavefunction. 

As we shall show, our NN-VMC uses a general architecture that is not at all customized for the FQH problem, and no information about FQH physics is put in by hand during NN initialization and training. Despite being completely unbiased and having no prior knowledge of physics, our FNN is able to reach large system size and solve the FQH problem at both weak and strong LL mixing with {\it unprecedented} accuracy. The wavefunction learned by our FNN exhibits distinctive pair distribution and phase structure at short distance $<\ell_B$ beyond the Laughlin ansatz.

%
\section{III. Neural network wavefunction}

We explore accurate variational wavefunctions $\Psi(\mathbf{r}_1, ..., \mathbf{r}_N)$ in real space to study 2D Coulomb gas under magnetic fields without LL truncation. Specifically, we consider a general class of variational wavefunctions built from generalized Slater determinants. While the standard Slater determinant $\text{det}(\Phi_i(\mathbf{r}_j))$ is constructed from a set of single-particle orbitals $\Phi_i(\mathbf{r}_j)$, here we generalize $\Phi_i$ to a many-body orbital $\Phi_i(\mathbf{r}_j, \mathbf{r})$ which also depends on the positions of all particles $\mathbf{r} \equiv \{ \mathbf{r}_1, ..., \mathbf{r}_N \}$. We further require that  $\Phi_i$ are \emph{permutation equivariant}: 
\begin{eqnarray}
\Phi_i(\mathbf{r}_j, \mathbf{r} ) = \Phi_i(\mathbf{r}_j, P \mathbf{r})\,,
\end{eqnarray}
where $P$ interchanges two elements in $\mathbf{r}$. Slater determinants constructed from such generalized orbitals $\text{det}(\Phi_i(\mathbf{r}_j, \mathbf{r}))$ are guaranteed to be antisymmetric under particle exchange. 

Artificial neural networks such as PauliNet and FermiNet provide a powerful tool to generate an expressive ansatz for generalized orbitals with a large number of variational parameters. The expressivity of these networks hinges upon combining one electron features $\{\mathbf{r}_i, \lvert \mathbf{r}_i \lvert \}$ and two electron correlation features $\{\mathbf{r}_i - \mathbf{r}_j, \lvert \mathbf{r}_i - \mathbf{r}_j \lvert \}$ in a versatile way via connected high-dimensional hidden layers. On the other hand, Psiformer, a self-attention based FNN, only takes in one electron features and self-generates electron correlations in a more flexible way by combining three sets of one electron features (\emph{Key}, \emph{Query} and \emph{Value}) with all-to-all intra- and inter-layer connectivity and arbitrary weights. During the combination process, all electron indices are symmetrized over owing to a careful design of the network architecture (more details in SM). This allows the correlations to be effectively captured while preserving permutation equivariance.  

\begin{figure}[h!]
    \centering
   \includegraphics[width=0.4\textwidth]{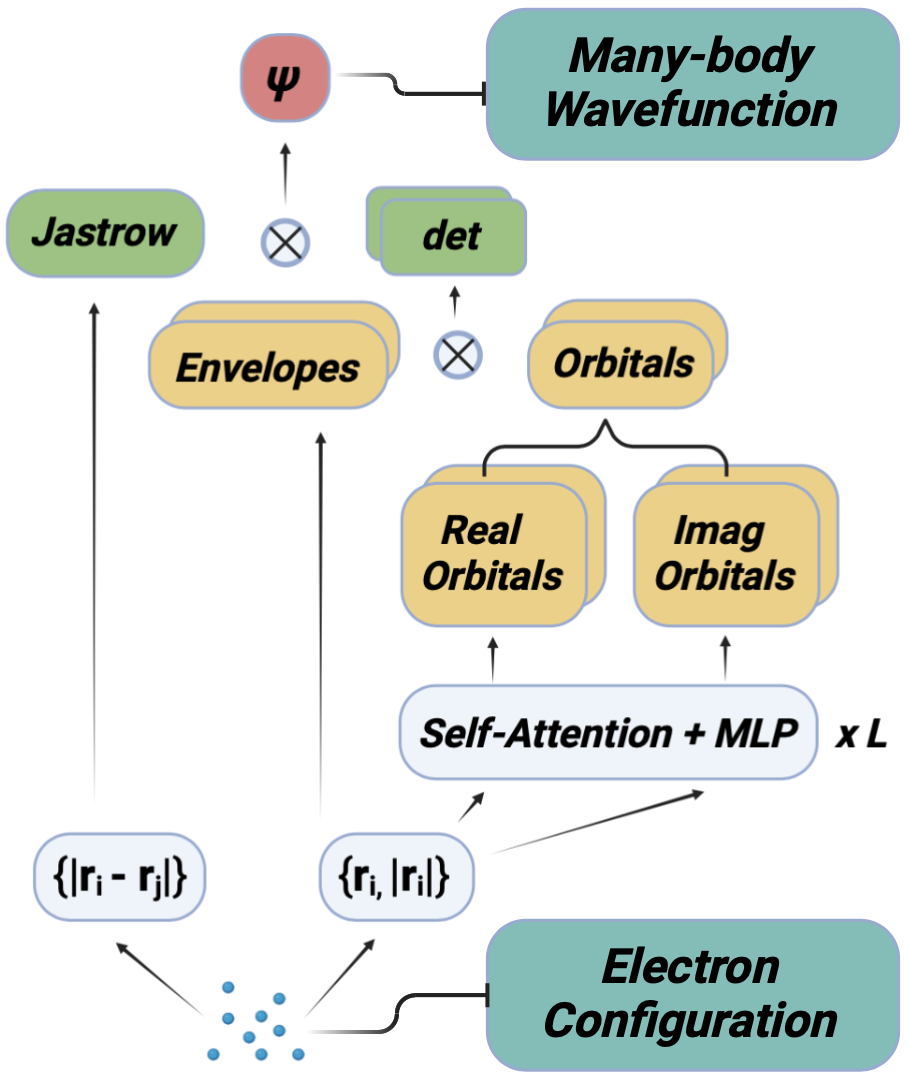}
    \caption{Psiformer architecture~\cite{VonGlehn_2023}: Electron configurations are used to generate one electron features and two electron features which are then fed into the network. One electron features are subsequently transformed through $L$ layers of self-attention and multi-layer perceptron (MLP) to form real and imaginary orbits. The complex orbits are then combined with isotropic Gaussian envelopes, after which a determinant and a Jastrow factor are applied to yield final many-body wavefunction $\Psi$.}
    \label{fig:fnn}
\end{figure}
In this work, we adapt Psiformer to 2D FQH system, with the network illustrated in Fig.~\ref{fig:fnn}. The real and imaginary parts of generalized Slater orbitals are created in separate streams via self-attention layers. An isotropic gaussian envelope $\pi_i \text{exp}(-\sigma_i r^2)$ is subsequently added to each orbital $\Phi_i$ to ensure the correct large distance behavior. These orbitals are then passed onto a determinant. Finally, to further enhance the expressivity, a sum of multiple determinants is used in combination with a Jastrow factor, yielding the final form of the many-body wavefunction
\begin{eqnarray}
\Psi_\theta(\mathbf{r})  &=& \text{exp}(\mathcal{J}_\theta(\mathbf{r})) \sum_{k = 1}^{N_{det}} \text{det}(\Phi_i^k(\mathbf{r}_j, \mathbf{r})) ,  \nonumber
   \\
   \nonumber
   \\
    \mathcal{J}_\theta(\mathbf{r}) &=& \sum_{\{i,j\}} - \beta \frac{\alpha^2}{\alpha+\lvert \mathbf{r}_i - \mathbf{r}_j\lvert},
\label{eq:fnn}
\end{eqnarray}
%
where $\boldsymbol{\theta}$ 
represents the set of learnable parameters. 
$\Phi_i^k(\mathbf{r}_j,\mathbf{r}))$ are permutation equivariant complex orbitals with its real and imaginary parts generated separately via self-attention layers (see Fig.~\ref{fig:fnn}). 
$\mathcal{J}(\mathbf{r})$ is the Jastrow factor with variational parameters $\alpha$ and $\beta$. Note that $\mathcal{J}$ takes real values and is to be distinguished from the Laughlin-Jastrow function $\prod_{ij} (z_i-z_j)^2$ in FQH. As we shall elaborate below, the Jastrow factor here plays a crucial role in capturing the behavior of the FQH wavefunction at short distance $<\ell_B$.  

\section{IV. Coulomb cusp and Jastrow factor}
Before proceeding, we shed some light on the role of the Jastrow factor $\mathcal{J}(\mathbf{r})$ and discuss the choice of initial value for the parameter $\beta$. For this purpose, we examine the short-distance behavior of the wavefunction where two electrons meet~\cite{Taut_2000, Foulkes_2001, Kralik_1997, Price_1996}. In general, the wavefunction of two spin-parallel electrons in two dimensions with relative angular momentum $l$ takes the following form at short distance: 
\begin{eqnarray}
   \psi(\rho, \theta, l) &=& e^{il \theta}\rho^{\lvert l \lvert} [1+\frac{1}{\epsilon (2\lvert l \lvert\, + 1) }\rho + \mathcal{O}(\rho^2)] \nonumber
   \\
   \nonumber
   \\
   &=& z^{\lvert l \lvert} [1+\frac{1}{\epsilon (2\lvert l \lvert\, + 1) }\lvert z \lvert+ \mathcal{O}(\lvert z \lvert^2)] 
\, , 
\label{eq:am_eigenfcn}
\end{eqnarray}
where ($\rho$, $\theta$) are the relative polar coordinates of two electrons, $z$ is the complex relative coordinate and $l$ is odd due to Fermi statistics and only takes positive values under high magnetic field. One important implication of Eq.~\eqref{eq:am_eigenfcn} is that the wavefunction is \emph{non-analytic} at $z \equiv z_1 - z_2 = 0$. The linear term $\sim \lvert z \lvert$ in the parenthesis has a discontinuity in derivative at $z = 0$ or equivalently $\mathbf{r}_1 = \mathbf{r}_2$, known as the Coulomb cusp. The prefactor of this $\lvert z \lvert$ term is called the cusp parameter, which depends on the Coulomb interaction strength $1/\epsilon$ and the relative angular momentum $l$.

Importantly, the Coulomb cusp is a universal feature of Coulomb systems, which {\it cannot} be captured by any wavefunction within the LLL. To see this, we note that LLL wavefunctions take the general form  
\begin{equation}
\Psi_{\rm LLL} = f(z_1, ..., z_N) \,e^{-{\Sigma_i \lvert z_i \lvert^2}/{4\ell_B^2}}
\label{eq:wavefcn_lll}
\end{equation}
where $f$ is a holomorphic function of complex electron coordinates $z_i$. For any given configuration of $N-2$ electron coordinates $\{ z_{3}, ..., z_N \}$, $f$ is a holomorphic function of the remaining two electrons' coordinates $z_{1, 2}$, and thus also a holomorphic function of the relative coordinate $z_{12} \equiv z_1-z_2$ and the center of mass coordinates $Z_{12} \equiv z_1 + z_2$. The Gaussian part can also be rewritten in terms of $z_{12}$ and $Z_{12}$ using $\lvert z_1 \lvert^2 +\, \lvert z_2 \lvert^2 \,= \lvert z_{12} \lvert^2/2 + \lvert Z_{12} \lvert^2/2$.
Combining the holomorphic part and the Gaussian part yields  
\begin{equation} 
\Psi_{\rm LLL} \sim g(z_{12}, Z_{12})\, e^{- \lvert z_{12} \lvert^2/{8\ell_B^2} } \,e^{ -\lvert Z_{12} \lvert^2/{8\ell_B^2}}\,,  \label{LLL}
\end{equation}
which cannot produce the linear dependence on $\lvert z_{12}\lvert$ and thus is incapable of capturing the Coulomb cusp shown in Eq.~\eqref{eq:am_eigenfcn}. In other words, LLL wavefunctions give qualitatively wrong results for pair correlation at ultra-short distance. In fact, in order to capture the {\it non-analytic} cusp feature, it is necessary to invoke the mixing of \emph{infinitely} many LLs.

This motivates the inclusion of a Jastrow factor as in Eq.~\eqref{eq:fnn}, which depends on the relative distance between two electrons $|\mathbf{r}_i - \mathbf{r}_j|$.  If the many-body wavefunction at short distance $|\mathbf{r}_i - \mathbf{r}_j|  < \ell_B$ is dominated by a particular relative angular momentum $l$ component, the aforementioned cusp condition dictates the choice of parameter $\beta = {1}/{\epsilon (2\lvert l \lvert\, + 1) }$. 
However, for FQH systems with Coulomb interaction, we expect a mixture of $l = 1$ and $l = 3$ components in the ground state wavefunction as two electrons approach each other: the $l=1$ component is the generic behavior of 2D Coulomb systems at ultrashort distance, while the success of Laughlin wavefunction suggests that the $l=3$ component should appear in $\nu=1/3$ FQH ground state at intermediate distance $\sim \ell_B$. To capture this crossover behavior, we choose the initial value for cusp parameter $\beta$ to be $1/4$, which lies between the value $1/3$ for $l=1$ and $1/7$ for $l=3$ (assuming $\epsilon=1$). This choice works well in practice, as shown by our results in section VI and VII, and further discussed in SM. 

\section{V. NN-VMC Simulation} 

\begin{figure}[t]
    \centering
    \includegraphics[width=0.48\textwidth]{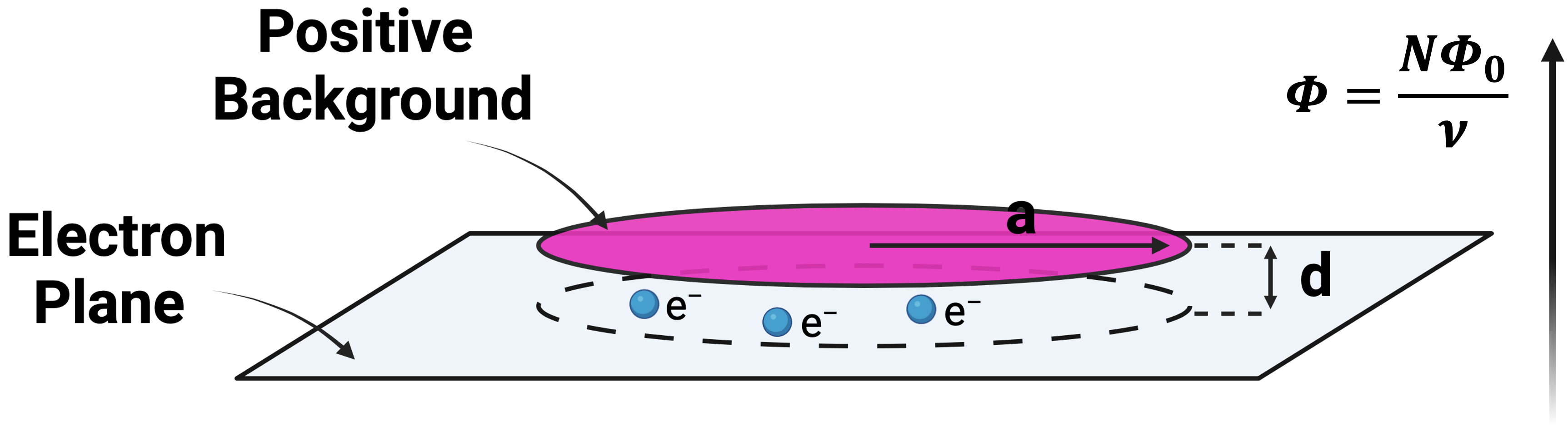}
    \caption{The geometry of our system: an infinite plane with $N$ electrons and a positively charged disk of radius $a$ at distance $d$ above. A total of $+Ne$ background charges is uniformly distributed and held fixed over the disk. A uniform magnetic field is applied perpendicular to the plane with the total flux through the disk being $\Phi = N\Phi_0/\nu$. }
    \label{fig:schematics}
\end{figure}

As shown in Fig.~\ref{fig:schematics}, our setup is $N$ spin-polarized electrons on an infinite 2D plane with Coulomb interaction. Parallel to and at distance $d$ above the plane is a uniformly charged disk of radius $a$ with a total charge $+Ne$, which provides a neutralizing background and a realistic confining potential. Moreover, a uniform magnetic field $B$ is applied perpendicular to the plane. 

The Hamiltonian of our system is given by 
\begin{eqnarray}
H &=&  H_0 + \sum_{j = 1}^{N} V_c(\mathbf{r}_j) + V_b \,,  \nonumber
   \nonumber
   \\
H_0 &=&  \sum_{j = 1}^{N} \frac{1}{2}(-i\boldsymbol{\nabla}_j + \frac{1}{2} \boldsymbol{B} \times \mathbf{r}_j )^2 + \sum_{i \neq j} \frac{1}{2\epsilon}\frac{1}{\lvert \mathbf{r}_i - \mathbf{r}_j \lvert}, 
\label{eq:hamiltonian}
\end{eqnarray}
where atomic units, namely $\hbar = e = m_e = 4\pi\epsilon_0 = 1$, are used, $\epsilon$ is the relative dielectric constant, and $V_c$ and $V_b$ are the confining potential and the background self-interaction energy. The details on confining potential and its implementation are given in the SM, and the constant background self-interaction is given by $+ 8 N^2/3 \pi \epsilon a$~\cite{Ciftja_2017}. 

Up to the rescaling of energy and coordinates, the Hamiltonian of our $N$-electron system is characterized by a single dimensionless quantity --- the LL mixing parameter $\lambda \equiv ({e^2}/{4\pi\epsilon_0\epsilon \ell_B})/\hbar\omega_c =  1/\epsilon\,\sqrt{\omega_c}$, where in the last equality we have converted from SI to atomic units. For a given material, while the effective electron/hole mass $m$ and dielectric constant $\epsilon$ are fixed, $\lambda \propto 1/\sqrt{B}$ varies with the applied magnetic field. 

The confining potential $V_c(\mathbf{r})$ consists of a flat region in the middle and a steep rise towards the edge ($r \rightarrow a$). For small disk separation $d \ll a$, the potential strongly traps the electrons to a disk of radius $a$, resulting in a well-defined effective area of $\pi a^2$. The disk radius $a$ is then set to have a total magnetic flux $\Phi = N\,\Phi_0/\nu$ that corresponds to Landau level filling $\nu$.  
All simulations are performed at $\nu=1/3$. 
We choose $d = 0.1 \ell_B$ so that no edge reconstruction occurs~\cite{Wan_2002, Wan_2003}. 


%

To perform NN-VMC simulation, we take the total energy of trial FNN wavefunction as our loss function, evaluate the loss function and its gradient and minimize it using the KFAC algorithm~\cite{Martens_2020_KFAC}, an approximate method to natural gradient descent~\cite{Stokes_2020}. The total energy is obtained by calculating the expectation value of the Hamiltonian:
\begin{equation}
\langle \hat{H} \rangle = \frac{\int \Psi_{\theta}^{*}(\mathbf{r})\hat{H} \Psi_{\theta}(\mathbf{r}) d\mathbf{r}^2}{\int \Psi_{\theta}^{*}(\mathbf{r})\Psi_{\theta}(\mathbf{r}) d\mathbf{r}^2} = \frac{\int \lvert \Psi_{\theta}(\mathbf{r}) \lvert^2 {E}_L(\mathbf{r})\,d\mathbf{r}^2}{\int \lvert \Psi_{\theta}(\mathbf{r}) \lvert^2 d\mathbf{r}^2}\,,
\label{eq:loss}
\end{equation}
where $\Psi_{\theta}(\mathbf{r})$ is the unnormalized variational wavefunction that depends on learnable parameters $\boldsymbol{\theta}$ and electron positions $\mathbf{r}$, and $E_L(\mathbf{r}) = \Psi_{\theta}^{-1}(\mathbf{r}) \hat{H}\Psi_{\theta}(\mathbf{r})$ is the local energy. Thus, we can estimate the total energy by sampling local energy according to a probability distribution $\lvert \Psi_{\theta}(\mathbf{r}) \lvert^2 $ proportional to electron density. The sampling is thus equivalent to simply averaging over the local energies obtained from different Monte Carlo sampling configurations. Subsequently, the gradient of the total energy is evaluated, and the network parameters are updated by KFAC optimizer.
\begin{figure*}[t]
    \includegraphics[width=0.98\textwidth]{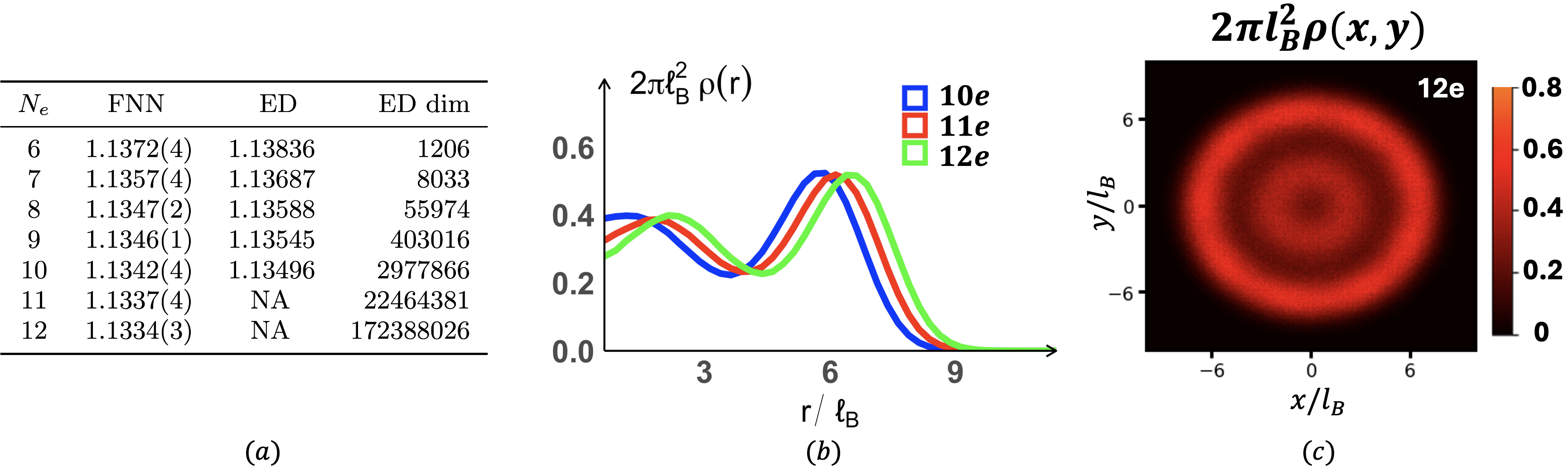}
    \caption{ED and FNN benchmark. (a) The ground state energy per particle (in the unit of $1/\epsilon \ell_B$) obtained from ED and FNN for LL mixing parameter $\lambda = 1/3$. ED is done up to 10 electrons due to limited resources. The dimension of LLL subspace in ED is given on the right.
    (b) The 10e, 11e and 12e spatial density sampled from FNN. 
    (c) 2D charge density for 12 electrons.} 
    \label{fig:results}
\end{figure*}

In VMC simulations, one usually looks for the convergence of the total energy and the vanishing variance of the local energy. The vanishing variance of the local energy indicates the resulting state being an energy eigenstate, while the convergence of the total energy further increases the confidence in the state being the ground state. However, this approach becomes less effective when the system has low-energy excited states. This is indeed expected to the case for our problem, as quantum Hall fluids support gapless edge excitations~\cite{Halperin_1982, Wen_1990}. Then, the training of neural network is often stuck at a superposition of low-lying states, in which case the total energy will appear to converge 
and the local energy fluctuation becomes nearly vanishing before actually reaching the true ground state. 

In order to overcome this difficulty and attain the fully converged results, we choose to measure the total angular momentum $M$ besides the local energy and total energy during our NN training. As our system is circularly symmetric, the total angular momentum of the ground state takes integer values, for example, 
$M_L = {m N(N-1)}/{2}$ for the Laughlin wavefunction of  $N$ electrons at filling $m = 1/\nu$. 
As a {\it discrete and quantized} quantity, the angular momentum is a clear indicator of how close the present trial wavefunction is to the actual ground state, because low-lying states often belong to different angular momentum sectors and can thus be distinguished. Therefore, monitoring the approachment of angular momentum to an integer provides a powerful way to check the convergence of NN-VMC. The details of angular momentum measurement are given in SM.
%
%

\section{VI. Results}

We first benchmark our FNN performance against LL-projected ED for 6-12 electrons in weak LL-mixing regime ($\lambda = 1/3$). The ED calculation is performed in the full subspace of all possible $N$-electron states within the LLL that have the total angular momentum $M_L$, that of the $\nu=1/3$ Laughlin wavefunction. As such, our ED energies are guaranteed to be lower than previous studies that restrict the total number of lowest Landau level orbitals accessible to electrons~\cite{tsiper_2001}.  The Hilbert space dimension of our ED grows rapidly with $N$, as shown under ED dimension in Fig.~\ref{fig:results}(a). Due to the memory limit, our ED is performed up to 10 electrons.

The ground state energies from FNN and ED are both given in Fig.~\ref{fig:results}(a). For a weak LL mixing parameter $\lambda=1/3$, our FNN consistently reaches energies slightly lower than ED, demonstrating its capacity to learn fractional quantum Hall ground states to higher accuracy. The fact that our variational wavefunction has lower energy than LL-projected ED demonstrates the presence of LL mixing.

The spatial profile of charge densities for 10, 11 and 12 electrons are plotted in Fig.~\ref{fig:results}(b), in which the radial charge modulation and the rise near the edge are consistent with previous ED studies in the disk geometry~\cite{tsiper_2001, Wan_2002, Wan_2003}. We also see that the charge density profile  already shows a trend of convergence even with 10-12 electrons, and one can expect the interior region further flattens in the thermodynamic limit. 

We also plot the 2D charge density for the 12 electrons in Fig.~\ref{fig:results}(c). Although we do not impose any circular symmetry into the wavefunction ansatz, the self-attention NN remarkably learns 
a ciruclarly symmetric ground state. This circular symmetry is further evidenced by the convergence of angular momentum to $197.97$, which is extremely close to the quantized value $M_L=198$ for the Laughlin state with 12 electrons. As a note in passing, we also found that the FermiNet has difficulties in attaining the correct angular momentum and hence the circular symmetry even for 9-10 electrons. More training details and additional training data are given in SM.

\begin{figure*}[!t]
    \includegraphics[width=0.98\textwidth]{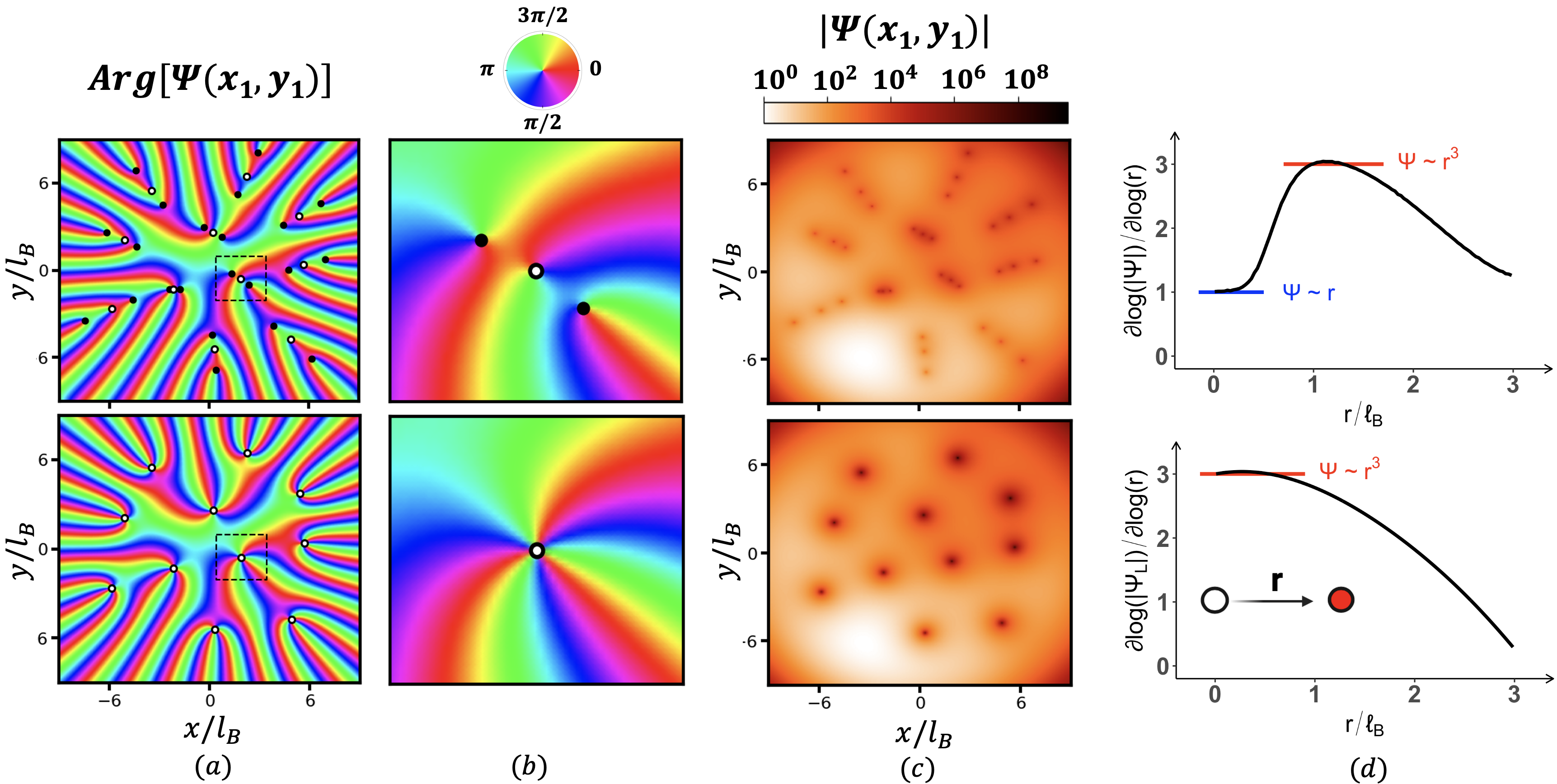}
    \caption{Visualization of the many-body wavefunctions for Coulomb interaction (top panel) and for the $\nu=1/3$ Laughlin state (bottom panel). The Landau level mixing parameter is chosen to be $\lambda = 1/3$ for the Coulomb interaction case. 
    (a) The phase of $N=12$ electron wavefunction $\text{Arg} [\Psi]$. Empty circles mark the $N-1$ fixed electrons. Solid circles mark the additional first-order zeros with $2\pi$ phase winding.
    (b) Zoom-in image of the dashed rectangular region marked in (a).  
    (c) The magnitude of the wavefunction $\lvert \Psi \lvert$ as a function of first electron coordinates with the remaining 11 electrons fixed. 
    (d) $\partial \text{log}(\lvert\Psi\lvert) / \partial \text{log}(r)$ reveals the scaling behavior of the wavefunction in terms of the two-electron relative distance.}
    \label{fig:visualization}
\end{figure*}

After training and learning the many-body wavefunction, our FNN allows for efficient computation of various ground state properties. We use the trained network to study microscopic features of the 12 electron Coulomb ground state (top panel in Fig.~\ref{fig:visualization}) and compare them with the $1/3$ Laughlin wavefunction (bottom panel). 
To probe electron correlation, 
we take $N-1$ electron positions $\mathbf{r}_{j\geq 2}$ (displayed as empty circles in Fig.~\ref{fig:visualization}(a)) from a typical Monte Carlo configuration (defined as the most probable configuration out of 5000 samples) and hold them fixed, and study the ground state wavefunction $\Psi(\mathbf{r}_1, ..., \mathbf{r}_N)$ as the remaining electron $\mathbf{r}_1$ moves through the 2D plane acting. 

The phase and magnitude of the wavefunction as a function of the first electron's coordinates $(x_1, y_1)$ are shown in Fig.~\ref{fig:visualization}. In constrast to Laughlin wavefunction which only shows third or higher order zeros when two electrons coincide, our NN wavefunction for the Coulomb ground state exhibits three first-order zeros when the electron $\mathbf{r}_1$ approaches another electron: one zero occurs at the intersection (required by Pauli exclusion principle) and two additional zeros nearby, as shown in Fig.~\ref{fig:visualization}. The phase winding number in the neighborhood of each zero is $l=1$ for Coulomb, so that the total winding number as one electron moves around the other at intermediate distance  $\sim \ell_B$ is $3$, recovering the behavior of Laughlin wavefunction.  

The splitting of each third-order zero into a cluster of three first-order zeros in Coulomb ground state can be understood from the mixing of $l = 1$ and $l = 3$ relative angular momentum components as in Eq.\ref{eq:am_eigenfcn}: 
\begin{eqnarray}
   \Psi(r, \theta) \sim A\, r e^{i \theta} +\, B\, r^3 e^{3i\theta} 
= A\,z_{12} + B\,z_{12}^3  
   \nonumber
   \\
  \sim (z_1 - z_2) \, [z_1 - (z_2 - \xi)][z_1 - (z_2 + \xi)]\,,
\label{eq:wavefcn_13mix}
\end{eqnarray}
where $r$, $\theta$ are relative polar coordinates and $z_{12}$ are relative complex coordinates of two nearby electrons. This short-distance behavior implies that the two additional zeros located at $z_1 = z_2 \pm \xi $ should align symmetrically with the original zero at $z_{1}=z_2$, so that the three zeros form a linear trimer. Indeed, these linear trimers of zeros are clearly observed in our FNN wavefunction, as shown in Fig.~\ref{fig:visualization}(b). 

The mixing of $l=1$ and $l=3$ components also leads to a crossover from $r$ to $r^3$ dependence of $\Psi$ as the relative distance $r$ between two electrons increases. To probe this behavior, we take the derivative of $\text{log}(\lvert\Psi\lvert)$ with respect to $\text{log}(r)$. 
As shown in Fig.~\ref{fig:visualization}(c), for Coulomb interaction (top panel), $\Psi \sim r$ as $r \rightarrow 0$ and $\Psi \sim r^3$ around $r = \ell_B$, whereas for Laughlin (bottom panel), $\Psi_L \sim r^3$ down to $r = 0$.

\section{VII. Landau level mixing and Quantum phase transition}

\begin{figure*}[!t]
    \includegraphics[width=0.98\textwidth]{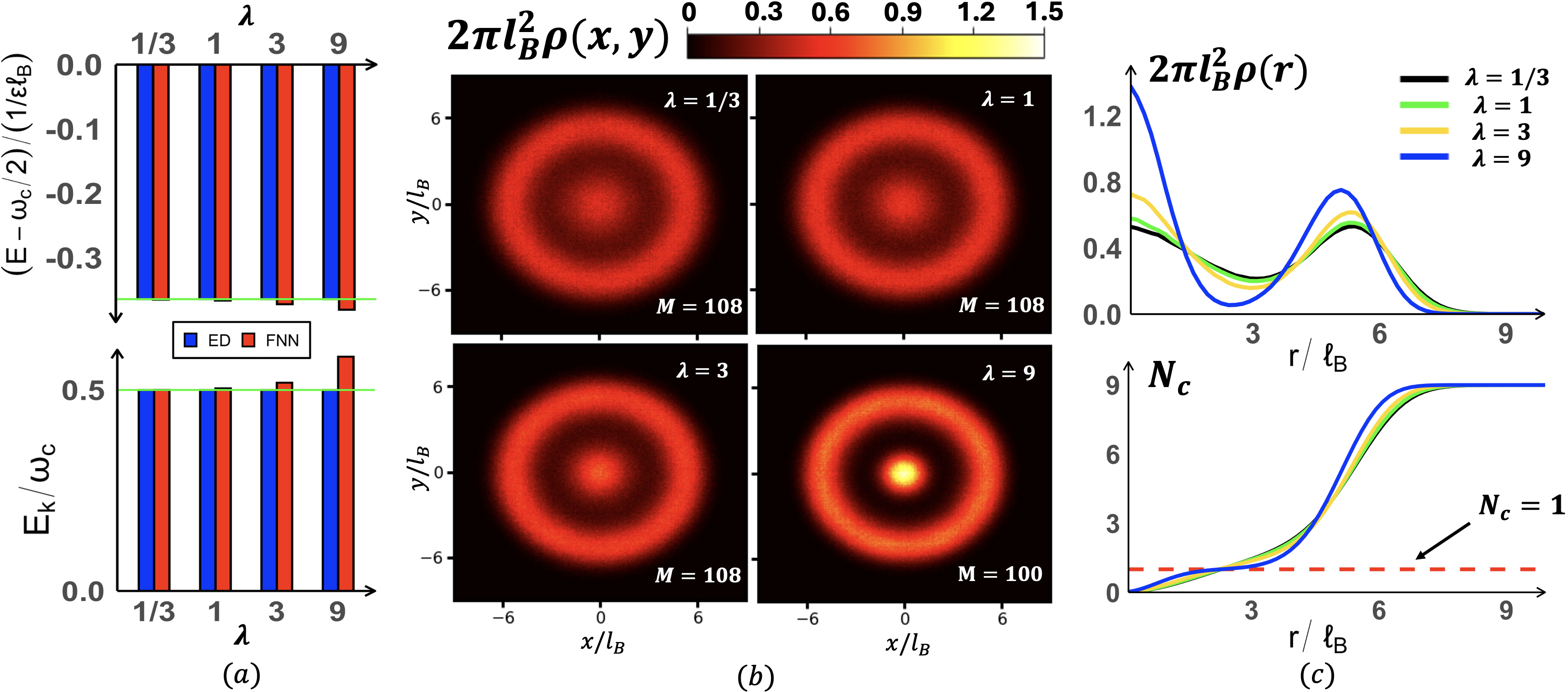}
  \caption{The results for 9 electrons at various Landau mixing parameters.
    (a) Top panel shows the ground state reduced energy per particle, defined as the total energy $E$ subtracted by $\omega_c / 2$ (the kinetic energy in LLL). The resulting reduced energy is plotted in the unit of $1/\epsilon \ell_B$. 
    Bottom panel shows the kinetic energy per particle in unit of cyclotron frequency $\omega_c$. It is equal to $1/2$ for LL-projected ED irrespective of interaction strengths.
    (b) The 2D real space charge density. Total angular momentum is given by $M = 108$ for $\lambda = 1/3,1$ and $3$, and $M = 100$ for $\lambda = 9$.
    (c) The charge density and cumulative charge density in radial direction are shown on the top and bottom panels respectively.} 
    \label{fig:ll_mixing}
\end{figure*}

We now proceed to explore the effects of increasing LL mixing parameter $\lambda$. We first measure the kinetic energy per particle (see Fig.~\ref{fig:ll_mixing}(a) bottom panel), which stays at $\omega_c/2$ in LLL-projected ED. Any increments from $\omega_c/2$ implies finite occupation in higher LLs, so it serves as a clear indicator of LL mixing. We can see for $\lambda = 3$ and $\lambda = 9$, the kinetic energy significantly exceeds $\omega_c/2$, revealing the presence of prominent LL mixing. 

Despite the increase in kinetic energy, LL mixing greatly reduces the interaction energy, thus giving rise to a decrease in the ground state energy. This effect can be quantified by considering the reduced energy per particle, defined as $E - \omega_c/2$, namely the difference between $E$, the total energy per particle, and $\omega_c/2$, the kinetic energy per particle in the LLL. This reduced energy includes (1) the electron-electron interaction, (2) the electrostatic interactions involving the positively charged disk, and (3) additional kinetic energy from the occupation of higher LLs.  Since both (1) and (2) are proportional to the Coulomb interaction strength $1/\epsilon$, the reduced energy expressed in unit of $1/\epsilon \ell_B$ is constant in LLL-projected ED where (3) is absent. The reduced energies for $\lambda = \{1/3, 1, 3, 9\}$ are plotted in Fig.~\ref{fig:ll_mixing}(a) top panel. By  variationally solving the full Hamiltonian that includes all LLs, FNN consistently reaches lower energy than LL-projected ED. This energy difference becomes more pronounced for $\lambda \geq 1$, demonstrating the significant role of LL mixing in minimizing energy.

We further search for the signature of LL-mixing induced quantum phase transition. It is expected that in the strong interaction limit where $1/\epsilon \rightarrow \infty$, a transition from FQH liquid to a Wigner crystal occurs. The Wigner crystal state is an array of tightly localized electrons, with the localization length $\sim \sqrt{\ell_B a_0} \ll \ell_B$ ($a_0$ is Bohr radius). Thus, strong LL mixing is necessary to drive the system into the Wigner crystal. As $\lambda$ increases, the FQH liquid may undergo a direction transition into the Wigner crystal, or through an intermediate phase. 

This problem has previously been studied with fixed-phase DMC, which employs separate initial trial wavefunctions for the FQH liquid and Wigner crystal, optimizes the magnitude of the wavefunctions while keeping their phase factors $\exp{i \varphi(\mathbf{r}_1, ..., \mathbf{r}_N)}$ fixed, and finally compares their energies. In contrast, our FNN is not predicated any customized ansatz or restricted to any fixed phase structures. As such,
it provides an unbiased and unified way to learn and discover various phases of matter with higher accuracy.

First, we examine the change in 2D charge density with a range of LL mixing parameters. As seen in Fig.~\ref{fig:ll_mixing}(b), as LL-mixing becomes stronger, the charges are pushed towards the center and the edge as opposed to being relatively uniform in FQH liquid. To quantify this, we look at the spatial charge density profile and cumulative radial charge, illustrated in Fig.~\ref{fig:ll_mixing}(c) top and bottom panel, respectively. From spatial profile one can clearly see that for $\lambda = 9$ case, the middle region is largely-depleted with charges far away from one another. The cumulative charge distribution shows a plateau at $N_c = 1$ (not observed for $\lambda =1/3, 1, 3$), implying a charge configuration with one electron in the middle and eight on the edge. 

Furthermore, a direct evidence of phase transition lies in the abrupt change in the ground-state angular momentum from $M = 108$ (expected for FQH) in the case of $\lambda =1/3, 1, 3$, to $M = 100$ for $\lambda = 9$. We have also tested 7-11 electrons with $\lambda = 9$ and consistently observe drops in $M$ compared to that of FQH ground states. This reduction in total angular momentum marks the emergence of a new ground state that is distinct from the FQH liquid phase, because the latter has an angular momentum equal to that of the Laughlin wavefunction, or larger due to edge reconstruction. Therefore, we conclude that as $\lambda$ increases, LL mixing induces a quantum phase transition from the FQH droplet to a new state on the disk.

What is the nature of the $\lambda=9$ ground state in the thermodynamic limit? Before addressing this issue, we should note that on a finite-size disk the exact ground state must be an eigenstate of total angular momentum and have circular symmetric charge density. The charge order in Wigner crystal phase, which breaks rotational symmetry, only emerges in the thermodynamic limit, when different angular momentum sectors become  degenerate. Since numerical study on the disk is inevitably subject to finite size and boundary effect, it is difficult to draw a definite conclusion from the above NN-VMC studies. Nontheless, in the following we discuss possible scenarios in light of our results. 

A plausible scenario is that the ground state on the disk is a rotating Wigner molecule, a quantum superposition (cat state) of classical charge configurations that restores the broken circular symmetry. This picture is supported by comparing the charge density profile in FNN ground state with that the classical configuration. For $7 \leq N \leq 9$, the energetically favored classical configuration has one charge at the center and $N-1$ charges on the second shell \cite{Reimann_2002}. This pattern was indeed observed in our FNN ground state at $\lambda=9$ as discussed above. Moreover, for $N=10$ (see SM), our FNN finds 2 electrons in the first shell and 8 electrons in the second shell, also in agreement with the corresponding classical configuration. These evidences suggest that the ground state  in the thermodynamic limit may be a Wigner crystal that is adiabatically connected to the classical limit.     

Another possibility is a Hall crystal~\cite{Tesanovic_1989}, characterized by the quantized Hall conductance and charge-density wave order in coexistence. This state may appear at an intermediate range of LL mixing parameters as an intervening phase between FQH liquid and Wigner crystal. In particular, a Hall crystal with fractionally quantized Hall conductance can be naturally realized by Bose condensation of low-lying magnetoroton excitations of FQH liquid, which have minimum energy at finite wavevector~\cite{Girvin_1986}. We leave a systematic study of various crystal phases induced by LL mixing to future work.

Before closing, we emphasize that the same NN architecture has been used to attain variational ground states of our system at various interaction strengths in a unified manner. This shows that NN is capable of not only solving  many-electron problems in continuous space with high accuracy, but also learning and discovering new phases of matter.


\section{VIII. Conclusion and outlook} 

In this work, we adapt a self-attention FNN ansatz (Psiformer) to solve a canonical many-electron problem: 2D Coulomb electron gas under magnetic fields. Using a general NN architecture that was originally developed for quantum chemistry with no \emph{a priori} knowledge of the underlying physics, we find FQH ground states over a range of interaction strengths $\lambda$ and reveal microscopic features in the many-body wavefunction beyond the Laughlin ansatz. Our FNN consistently attains energies lower than LLL-projected ED and reaches larger system size. The unprecedented power of our FNN is demonstrated by strong LL mixing in the many-body wavefunction at large $\lambda$ (e.g., $\lambda=3$), for which unbiased and accurate numerical study has been lacking until now.     

At very large LL mixing parameter, our FNN finds another type of ground state, whose density profile and angular momentum are markedly different from the FQH liquid. This state is consistent with a rotating Wigner crystal on a finite-size disk. Our findings of FQH liquids and Wigner crystals in 2D Coulomb gas under magnetic fields demonstrate the remarkable universality of NN-VMC in discovering fractionalized and correlated phases of matter with a unifying NN architecture in an unbiased manner. One cannot help but marvel at the power of a single FNN architecture in accurately solving the ground states of atoms, molecules, solids, and now fractionalized quantum matter!    

Looking forward, our work invites further developments of FNN to address longstanding open problems in the field of fractional quantum Hall effect, and in particular problems related to strong LL mixing. These include quantum Hall liquid to Wigner crystal transition, Hall crystals, and non-Abelian even-denominator states. Even more exciting is the prospect of using FNN to discover new phases of matter in moir\'e materials, where numerous fascinating quantum phenomena related to topological band, strong correlation, symmetry breaking and electron fractionalization are being uncovered.    

{\it Acknowledgement} - 
We thank Xiang Li for generously sharing his expertise on fermionic neural network. We are grateful for GPU computing time provided by AI Tennessee Initiative seed fund to Yang Zhang as well as Max Planck Institute and MIT Office of Research Computing and Data. We thank Liangzi Yao, Bennet Becker, Hao-ren You, Michael Rutter and Christopher Hill for software support; Andres Perez Fadon and Matthew Foulkes for drawing our attention to cusp conditions; David Pfau, Tzen Ong, Chun-Tse Li and Kun Yang for helpful discussions. This work was supported by a Simons Investigator Award from the Simons Foundation. YT acknowledges support from MPIPKS Visitors Program.  
DDD was supported by the Undergraduate Research Opportunities Program at MIT. 
LF acknowledges support from the NSF through Award No. PHY-2425180. 

\bibliographystyle{apsrev4-2}
\bibliography{reference.bib}

\clearpage
\newpage

\appendix

\setcounter{figure}{0}
\renewcommand{\thefigure}{S\arabic{figure}}
\renewcommand{\appendixname}{SM Note}

\onecolumngrid
\vspace{\columnsep}
\begin{center}
{\Large\bf Supplementary Material} 
\end{center}

\section{\large Detailed Architecture of Psiformer \label{app:fnn_details}}

Here we mostly follow the original Psiformer paper~\cite{VonGlehn_2023}. As an ab-initio Ansatz, we start with only one-electron and two-electron features: $\{\mathbf{r}_i, \lvert \mathbf{r}_i \lvert\}$ and $\{\mathbf{r}_{ij}\}$. If we choose to rescale the input, then $\{\mathbf{\bar{r}}_i, \bar{r}_i\} = \{\mathbf{r}_i\frac{\text{log}(1+\lvert \mathbf{r}_i \lvert)}{\lvert \mathbf{r}_i \lvert}, \text{log}(1+\lvert \mathbf{r}_i \lvert) \} $ will be used so that the one-electron features grow logarithmically with distance. As we will see, this rescaling is not relevant for two-electron features as they only appear via a fixed functional form in the Jastrow factor. 

Initially, $\{\mathbf{r}_i, \lvert \mathbf{r}_i \lvert\}$ are concatenated as ${\boldsymbol{f}_i^0}$ and undergo a linear transformation (LT): ${\boldsymbol{h}_i^0} = {\boldsymbol{W}^0 \boldsymbol{f}_i^0}$. Then this ${\boldsymbol{h}_i^0}$ is transformed through several layers of self-attention and multi-layer perceptions (MLPs). Now we explicitly spell out the mathematical expressions going from layer L to L+1. First, several sets of \{key, query and value\} are created from the same ${\boldsymbol{h}_i^L}$ but with different learnable LT matrices $\boldsymbol{W}_{k,q,v}^{L\alpha}$
\begin{equation}
   \boldsymbol{k}_i^{L\alpha} =  \boldsymbol{W}_k^{L\alpha} \boldsymbol{h}_i^L, 
   \boldsymbol{q}_i^{L\alpha} =  \boldsymbol{W}_q^{L\alpha} \boldsymbol{h}_i^L, 
   \boldsymbol{v}_i^{L\alpha} =  \boldsymbol{W}_v^{L\alpha} \boldsymbol{h}_i^L,
\label{eq:fnn_kqv}\tag{S1}
\end{equation}
where $\alpha$ labels different attention heads (different sets of \{key, query and value\}), and i labels different neurons. Then the correlations between the keys and queries of all neurons are computed in the form of dot product. These correlations are subsequently used as weights to sum the value set $v_i^{L\alpha}$ and produce the update matrices $U_i^{L\alpha}$.
\begin{equation}
    \boldsymbol{U}_i^{L\alpha} = \frac{1}{\sqrt{d}}\sum_j \sigma_j(\boldsymbol{q}_1^T \boldsymbol{k}_i, ..., \boldsymbol{q}_N^T \boldsymbol{k}_i) v_j
\label{eq:fnn_update}\tag{S2}
\end{equation}
where $\sigma_i(\mathbf{r}_1,...,\mathbf{r}_N) = \frac{\text{exp}(\mathbf{r}_i)}{\sum_j \text{exp}(\mathbf{r}_j)}$ is the softmax function, and d is the dimension of key/query output. Finally, we concatenate and activate the attention update matrices as follows 
\begin{equation}
    \boldsymbol{f}_i^{L+1} = \boldsymbol{h}_i^L + \boldsymbol{W}^L\,\text{concat}[\boldsymbol{U}_i^{L1}, ..., \boldsymbol{U}_i^{Lm}], \,\,
    \boldsymbol{h}_i^{L+1} = \boldsymbol{f}_i^{L+1} + \text{tanh}(\boldsymbol{V}^{L+1} \boldsymbol{f}_i^{L+1} + \boldsymbol{b}^{L+1}),
\label{eq:fnn_fh}\tag{S3}
\end{equation}
where $\boldsymbol{U}_i^{L\alpha}$ are attention outputs from different attention heads, and m denotes the total number of heads. The next step is to combine the envelopes and include both real and imaginary parts to arrive at the complex-valued orbitals: 
\begin{equation}
    \Phi_{ij} = \Omega_{ij}\,\boldsymbol{M}_i^T \boldsymbol{h}_j^{L_f}
, \Phi_{ij} \rightarrow \Phi_{ij}^{real} + i \,\Phi_{ij}^{imag}, 
\label{eq:fnn_orbital}\tag{S4}
\end{equation}
where $\Omega_{ij} = \pi_i \text{exp}(-\sigma_i\lvert r_j \lvert^2)$ represents the isotropic gaussian envelope, $\boldsymbol{M}$ is learnable. Finally, we take determinant, linearly combine multiple determinants and multiply by the desired Jastrow factor,
\begin{equation}
    \Psi_\theta(\mathbf{r})  = \text{exp}(\mathcal{J}_\theta(\mathbf{r})) \sum_{k = 1}^{N_{\text{det}}}\text{det}(\Phi_{real}^{k} + i\Phi_{imag}^{k}), 
\label{eq:fnn_wavefcn}\tag{S5}
\end{equation}
where the Jastrow factor~\cite{Ceperley_1978, Foulkes_2001} is given by 
\begin{equation}
    \mathcal{J}_\theta(\mathbf{r}) = \sum_{\{i,j\}} - \beta \frac{\alpha^2}{\alpha+\lvert \mathbf{r}_i - \mathbf{r}_j\lvert},
\label{eq:fnn_jastrow}\tag{S6}
\end{equation}
and $\alpha$ is a learnable parameter. The choice of $\beta$ depends on spin-parallel/anti-parallel, dimension of the system and the asymptotic forms of wavefunction when two electrons approach each other. It is discussed in great details in the next part.

%
%
\section{\large Jastrow Factor and Cusp Conditions \label{app:jastrow}}
The motivation for a Jastrow factor is that under Coulomb interaction, there is a cusp (discontinuity) in the derivatives of the wavefunction at the point where two charged particles meet. This is discussed in details in \cite{Ceperley_1978, Ceperley_2016, Foulkes_2001}. Here we focus on the spin-parallel electron-electron cusps in two dimensions. Consider the two electron with coordinates $\mathbf{r}_1, \mathbf{r}_2$ defined by the following hamiltonian:
\begin{equation}
H = \frac{1}{2}[-i\boldsymbol{\nabla}_1 + \boldsymbol{A}(\mathbf{r}_1)]^2 + \frac{1}{2}[-i\boldsymbol{\nabla}_2 + \boldsymbol{A}(\mathbf{r}_2)]^2 + \frac{1}{\epsilon\lvert \mathbf{r_1} - \mathbf{r_2} \lvert} + V_c(\lvert\mathbf{r}_1\lvert) + V_c(\lvert\mathbf{r}_2\lvert),
\label{eq:two_body_ham}\tag{S7}
\end{equation}
where $\boldsymbol{\nabla}_1, \boldsymbol{\nabla}_2$ act on the coordinates $\mathbf{r}_1, \mathbf{r}_2$, $V_c$ is the confining background potential, and $\epsilon$ is the relative dielectric constant. In our case, the two electrons sit in 2D uniform magnetic field with a confining potential that is everywhere finite. Transforming the coordinates into center of mass coordinate $\mathbf{R} = (\mathbf{r}_1 + \mathbf{r}_2) / 2$ and relative coordinate $\mathbf{r} = \mathbf{r}_1 - \mathbf{r}_2$, and using symmetric gauge $A = (-By, Bx, 0)/2$, we arrive at 
\begin{equation}
H = - \frac{1}{4} \boldsymbol{\nabla}_{\mathbf{r}}^2 -
 \frac{i}{2} \boldsymbol{B} \cdot (\mathbf{r} \wedge \frac{\partial}{\partial \mathbf{r}}) + \frac{1}{16} B^2 r^2 -\boldsymbol{\nabla}_{\mathbf{R}}^2 - \frac{i}{2} \boldsymbol{B} \cdot (\mathbf{R} \wedge \frac{\partial}{\partial \mathbf{R}}) + \frac{1}{4} B^2 R^2 +
 \frac{1}{\epsilon\lvert \mathbf{r} \lvert} + V_c(\lvert\mathbf{r}+\mathbf{r}/2\lvert) + V_c(\lvert\mathbf{r}-\mathbf{r}/2\lvert).
\label{eq:two_body_ham_simplified}\tag{S8}
\end{equation}
For an energy eigenstate, the local energy $E_L = \psi^{-1} H \psi$ is constant everywhere, so the divergences in $E_L$ in the limit $\lvert r \lvert \rightarrow 0$ must cancel out each other. All the terms involving center of mass coordinates are clearly finite as $r \rightarrow 0$ and can be safely discarded. A neat way to proceed is to simply set $\mathbf{R} = 0$. We further assume that the derivatives of the confining potentials are well-behaved, then hamiltonian is then simplified into
\begin{equation}
H = - \boldsymbol{\nabla}_{\mathbf{r}}^2 - \frac{i}{2} \omega_c \frac{\partial}{\partial \theta} +
 \frac{1}{\epsilon\lvert \mathbf{r} \lvert} + [\frac{1}{16} \omega_c^2 + \frac{1}{4} V_c^{(2)}(0) ]\, r^2 + V_c(0) + \mathcal{O}(r^3)
\label{eq:two_body_ham_small_dis}\tag{S9}
\end{equation}
where $V_c(0)$ and $V_c^{(2)}(0)$ are $V_c$ and its second derivative evaluated at $r = 0$. We now ignore the quadratic term $\sim V_c^{(2)} r^2$ under the assumption that the potential does not vary much on the scale of magnetic length $l_B$. Since the hamiltonian commutes with $L_z = - i {\partial}/{\partial \theta}$, the z-component of relative angular momentum is a good quantum number (labeled $l$). Using separation of variables $\psi_l(r,\theta) = e^{il\theta} u(r)$ ($l \in \mathbb{Z}$), we obtain the 2d schrödinger equation of the form $\psi^{-1} H \psi = E$:
\begin{equation}
 - \frac{1}{u r}\frac{d}{d r}(r \frac{d u}{d r}) - \frac{l^2}{r^2} + \frac{1}{2} \omega_c l + \frac{1}{\epsilon\lvert \mathbf{r} \lvert} + \frac{1}{16}\omega_c^2 r^2 + V_c(0) = E.
\label{eq:two_body_schrodinger}\tag{S10}
\end{equation}

Rearranging the equation and absorbing all constants into the energy term, we arrive at
\begin{equation}
 \frac{d^2 u}{d r^2} + \frac{1}{r}\frac{d u}{d r}   + (\frac{l^2}{r^2} - \frac{1}{\epsilon\lvert \mathbf{r} \lvert} - \frac{1}{16}\omega_c^2 r^2 + \tilde{E}) u = 0, 
\label{eq:two_body_schrodinger_arranged}\tag{S11} 
\end{equation}
\begin{gather*}
  \tilde{E} = E - \frac{1}{2} \omega_c l - V_c(0).
\end{gather*}

Following an approach similar to 2d hydrogen atom case, we first transform energies and coordinates into dimensionless quantities by $\tilde{E} = - {1}/{\epsilon^2 \delta^2}$ and $r = \delta \epsilon x$. Using the ansatz $u(x) = x^{\lvert l \lvert} f(x)$, one can derive the following ODE
\begin{equation}
 x \frac{d^2 f}{d x^2} + (2\lvert l \lvert\, + 1 - \frac{x}{2}) \frac{d f}{d x} - (\delta + \frac{x}{4} - \gamma x^3) f(x)  = 0, \,\,\text{where} \,\, \gamma = \frac{1}{16} \omega_c^2 \epsilon^4 \alpha^4.
\label{eq:final_ode} \tag{S12}
\end{equation}

Though this does not admit close form finite series, we can make progress by series expansion
\begin{gather*}
  f(x) = 1 + a_1 x + \dots,
\end{gather*}

Keeping $\mathcal{O}(1)$ terms in Eq.\eqref{eq:final_ode}, we derive 
\begin{gather*}
  (2\lvert l \lvert\, + 1) a_1 - \delta  = 0,
\end{gather*}
\begin{equation}
   a_1 = \frac{\delta}{(2\lvert l \lvert\, + 1)}.
 \label{eq:a1_coeff} \tag{S13}
 \end{equation}

Note that though $\gamma$ did not appear explicitly in $a_1$, it alters the eigenenergies $E$ in a complicated way and hence determines the value(s) of allowed $\delta$. In general, this series would not terminate nicely as 2D hydrogen case, which seems to lead to two problems. First, the series might not converge for large x. This can be reconciled by the fact that in obtaining the final equation (Eq.~\eqref{eq:final_ode}), we have assumed small distance and hence no significant confining potential contribution. At large distance, we have a confining potential which effectively adds an exponential decaying envelope, regulating the behaviors of the wavefunction. Secondly, the linear coefficient appears to depend on an unknown quantity $\delta$. However, as we will see, the coefficient in front of the linear term remarkably becomes $\delta$ independent after transforming back to the original coordinate $r$
\begin{gather*}
f(r) = 1 + \frac{\delta}{(2\lvert l \lvert\, + 1)} \frac{r}{\epsilon \delta} + \cdots
 = 1 + \frac{1}{\epsilon (2\lvert l \lvert\, + 1) }r + \dots = C\, \text{exp}({- \frac{1}{\epsilon (2\lvert l \lvert\, + 1) } \frac{\alpha^2}{\alpha + \lvert \mathbf{r}_1 - \mathbf{r}_2 \lvert}}) + \mathcal{O}(r^2)
 \end{gather*} 
where C is a normalization constant. We see that the Jastrow factor in Psiformer is recovered with $\beta = {1}/{\epsilon (2\lvert l \lvert\, + 1)}$, where $m$ is the asymptotic relative angular momentum as two electrons approach each other.

It is instructive to step back and look closer into the final form of our wavefunction
\begin{equation}
   \psi(r, \theta) = e^{i l \theta} r^{\lvert l \lvert} [1 + \frac{1}{\epsilon (2\lvert l \lvert\, + 1) }r + \mathcal{O}(r^2)]
 \label{eq:final_wavefcn} \tag{S14}
 \end{equation}
where the relative angular momentum $l$ depends on both the magnetic field and the confining potential. In particular, in the high field limit, only positive $l$ components survive.

\section{\large FermiNet for Two Electron Cusp \label{app:ferminet_2e}}
Following the discussion in the last section, we proceed to consider the relevant cusp condition for 1/3 FQH states. We expect an interplay between $l = 1$ and  $l = 3$ components in Eq.~\eqref{eq:final_wavefcn}
\begin{equation}
   \psi(r, \theta) \sim A\, e^{i \theta} r [1 + \frac{1}{3}r + \mathcal{O}(r^2)] + B\, e^{3 i \theta} r^3 [1 + \frac{1}{7}r + \mathcal{O}(r^2)]
 \label{eq:wavefcn_13} \tag{S15}
 \end{equation}
where $\epsilon$ is set to 1 in our case, $A$, $B$ are constants. As expected from Laughlin theory and shown in previous studies, $B \gg A$, thus we cannot discard $l = 3$ component even down to very small distance ($\frac{1}{7} B\,r^4 \sim \frac{1}{3} A\, r^2$ when $B\,r^3 \sim A\,r$). Mathematically, the cusp at exactly $r = 0$ is $1/3$, but in practice we choose the cusp that well captures intermediate distance range, which means choosing a cusp parameter $\beta$ between $1/3$ and $1/7$.

\begin{figure*}[!t]
    \includegraphics[width=0.98\textwidth]{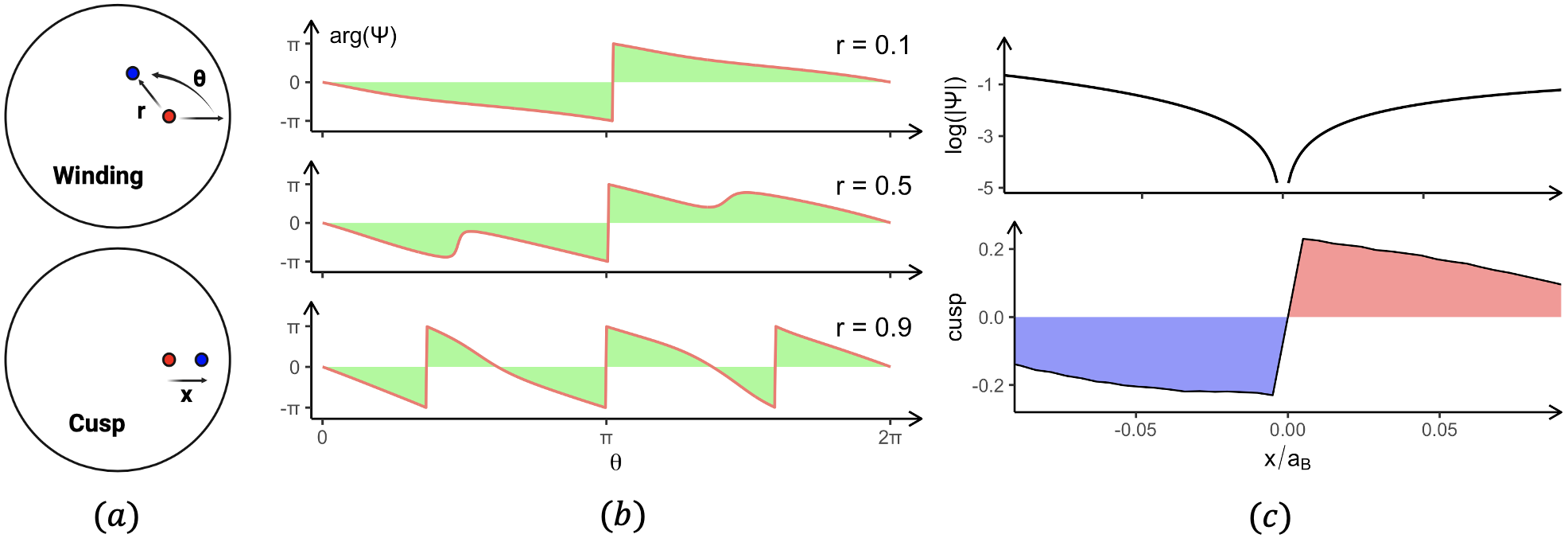}
    \caption{
    (a) Schematics of probing electron-electron cusps and winding. We fix one electron (red) at (0.6 $a_B$, 0) and move the other electron (blue) either around or across the fixed electrons, giving rise to winding phases or cusps, respectively.
    (b) The phase of the wavefunction as one electron moves around the other. Each green triangle corresponds to a phase shift of $\pi$. 
    (c) The logarithm of absolute value of wavefunction $\text{log}(\lvert \Psi \lvert)$ and the derivative of $\text{log}(\lvert \Psi \lvert / \lvert x \lvert)$ with respect to $x$. A constant offset has been added to both $x>0$ and $x<0$ due to some anisotropic linear dependence $\sim \boldsymbol{a}\cdot\mathbf{r}$. The original FermiNet architecture was used to calculate this case as it takes $\lvert r_1 - r_2 \lvert$ as input feature and can learn the cusp.}
    \label{fig:cusp_winding}
\end{figure*}

Before evaluating the cusp, we first verify the mixing between $l = 1$ and $ l =3$ by looking at the change in phase of wavefunction when one electron winds around the other, as shown in top panel in Fig.~\ref{fig:cusp_winding}(a). We fix one electron (red dot) to be at $(x,y) = (0.6 a_B, 0)$ (a point where the electron density is relatively high), and wind around the other electron (blue dot) at fix distance $r$. We see in Fig.~\ref{fig:cusp_winding} that as winding angle $\theta$ goes from 0 to $2\pi$, at very small distance $r = 0.1$ (top panel), the wavefunction phase also alters by $2\pi$ in a reasonably uniform fashion, which means $m = 1$ term dominates. At $r = 0.9$ (bottom panel), the wavefunction travels through $6\pi$ in total, corresponding to $m = 3$. In the intermediate region, some interference effect hints on a mix of $m = 1$ and $m = 3$. This also confirms that $B \gg A$ in Eq.~\eqref{eq:wavefcn_13} as $l = 3$ already dominates at $r = 0.9 < 1$.

To evaluate the cusp, we fix the first electron (red dot) on x-axis and moves the second electron (blue dot) along x-axis across the first one (bottom panel in Fig.~\ref{fig:cusp_winding}). We define the cusp parameter $\beta$ as 
\begin{equation}
\beta = \frac{1}{2}(\frac{\partial \text{log}(\lvert\Psi\lvert / \lvert x \lvert) }{\partial x}\bigg|_{0^{+}} - \frac{\partial \text{log}(\lvert\Psi\lvert / \lvert x \lvert) }{\partial x}\bigg|_{0^{-}})
\label{eq:cusp_condition}\tag{S16}
\end{equation}
where $x = x_2 - x_1$ is the x component of relative coordinates. This derivative is plotted in the bottom panel of Fig.~\ref{fig:cusp_winding}(c) up to a constant offset due to potential anisotropy. We can see that the appropriate $\beta$ to choose is indeed in between $1/3$ ($m=1$) and $1/7$ ($m=3$). In practice, we find $\beta \sim 0.25$ works well.
The cusp information learned from FermiNet can then be put into our more expressive Psiformer ansatz in a Jastrow factor of the following form:
\begin{equation}
    \mathcal{J}(\mathbf{r}) = \sum_{\{i,j\}} - \frac{1}{4} \frac{\alpha^2}{\alpha+\lvert \mathbf{r}_i - \mathbf{r}_j\lvert},
\label{eq:fnn_jastrow}\tag{S17}
\end{equation}
where $\alpha$ is a variational parameter. It is trivial to check that this Jastrow factor gives the correct small distance behavior as in Eq.~\eqref{eq:final_wavefcn}. Note that when $\epsilon \neq 1$, the Jastrow factor needs to be modified accordingly.

\section{\large JAX Implementation of the Confining Potential \label{app:potential_imp}}

The confining potential can be written in integral form:
\begin{equation}
V_c(\mathbf{r}) = - \frac{N}{\pi a^2} \int_{\lvert \mathbf{r'} \lvert < a} \frac{d\mathbf{r'}^2}{\sqrt{d^2 + \lvert \mathbf{r'} - \mathbf{r} \lvert^2}}
\label{eq:potential}\tag{S18}
\end{equation}
where d and a are disk separation and radius of the disk defined in Fig.\ref{fig:schematics} in main text.

We start with the double integral expression for the Coulomb potential:
\begin{equation}
V_c(\mathbf{r}) = - \frac{N}{\pi a^2} \int_{\lvert \mathbf{r'} \lvert < a} \frac{d\mathbf{r'}^2}{\sqrt{d^2 + \lvert \mathbf{r'} - \mathbf{r} \lvert^2}}
= - \frac{N}{\pi a^2} \int_{r' = 0}^{a} \int_{\theta = 0}^{2\pi}\frac{d\mathbf{r'}^2}{\sqrt{d^2 + r'^2 + r^2 - 2 r r' \text{cos}(\theta)}}
\label{eq:potential_double_int}\tag{S19}
\end{equation}

After integrating over $\theta$ using Mathematica, the remaining integral manifests circular symmetry and reads
\begin{equation}
V_c(r) = - \frac{N}{\pi a^2} \int_{r' = 0}^{a} r'\,(\,\frac{2 \sqrt{1 + \frac{4 r r'}{d^2 + (r - r')^2} }\, K[-\frac{4 r r'}{d^2 + (r + r')^2}]}{\sqrt{d^2 + (r + r')^2}} + \frac{2 \sqrt{1 - \frac{4 r r'}{d^2 + (r + r')^2} } \, K[\,\frac{4 r r'}{d^2 + (r - r')^2}]}{\sqrt{d^2 + (r - r')^2}}\,)\, dr',
\label{eq:potential_single_int}\tag{S20}
\end{equation}
where K(m) is the complete elliptic integral of first kind and is given by 
\begin{equation}
K(m) = \int_0^{\pi/2}(1 - m\, \text{sin}(t)^2)^{-1/2} dt\,.
\label{eq:ellipt_int}\tag{S21}
\end{equation}

To make the integral expression in Eqn.\eqref{eq:potential_single_int} compatible with gradient evaluation in JAX, we approximate the potential as follows:
\begin{equation}
  V(r)=\begin{cases}
    \text{\texttt{jax.interp}}\,(V_c), &  r<15a,\\
    -N \frac{1}{\sqrt{d^2 + r^2}}, & \text{otherwise}.
  \end{cases}
\label{eq:potential_double_int}\tag{S22}
\end{equation}
where \texttt{jax.interp} gives a linear approximation of $V_c$ given an array of its value pre-evaluated at a large number of different r (10000 points were used in our case), and at very large r ($r/a > 15$), we simply approximate the disk as a point charge. In practice, we never have any sample points in the region $r > 15a$ due to negligible density. We have compared the exact potential profile and the approximate profile, and have found the fractional error at all points are below 0.1\%. This protocol can in principle be extended to any potential without close form solutions.

Our last remark is on the convexity of the potential profile. It can be shown that the function is convex all the way up to $r = a$, beyond which lies negligible charge density. This means that the total potential energy evaluated using this method is necessarily a upper bound of the exact value, perfect for our benchmarking purposes.

\section{\large Angular Momentum Measurements \label{app:am_measurement}}
The expectation value of a general observable $\hat{O}$ from a unnormalized wavefunction $\Psi_{\theta}(\mathbf{r})$ can be defined in the following manner
\begin{equation}
\langle \hat{O} \rangle = \frac{\int \Psi_{\theta}^{*}(\mathbf{r})\hat{O} \Psi_{\theta}(\mathbf{r}) d\mathbf{r}^2}{\int \Psi_{\theta}^{*}(\mathbf{r})\Psi_{\theta}(\mathbf{r}) d\mathbf{r}^2} = \frac{\int \lvert \Psi_{\theta}(\mathbf{r}) \lvert^2 \,[\Psi_{\theta}^{-1}(\mathbf{r}) \hat{O}\Psi_{\theta}(\mathbf{r}) ]\,d\mathbf{r}^2}{\int \lvert \Psi_{\theta}(\mathbf{r}) \lvert^2 d\mathbf{r}^2} = \frac{\int \lvert \Psi_{\theta}(\mathbf{r}) \lvert^2 {O}_L(\mathbf{r})\,d\mathbf{r}^2}{\int \lvert \Psi_{\theta}(\mathbf{r}) \lvert^2 d\mathbf{r}^2} = \langle O_L \rangle
\label{eq:ops_exp} \tag{S23}
\end{equation}
where $O_L(\mathbf{r}) = \Psi_{\theta}^{-1}(\mathbf{r}) \hat{O}\Psi_{\theta}(\mathbf{r})$ is known as the 'local value' of the operator. We see that by factoring out $\lvert \Psi_{\theta}(\mathbf{r}) \lvert^2 $, the expression is transformed into sampling $O_L(\mathbf{r})$ over a unnormalized probability distribution $P_{\theta}(\mathbf{r}) = \lvert \Psi_{\theta}(\mathbf{r}) \lvert^2$. In contrast to the operator nature of $\hat{O}$, $O_L(\mathbf{r})$ is a function in position and can be readily evaluated and sampled. The evaluation is further simplified because the probability distribution coincides with the electron density distribution, so the expectation value of $O_L(\mathbf{r})$ is simply the average of $O_L(\mathbf{r})$ over all batch configurations (and over multiple MCMC steps when higher accuracy is needed). One celebrated example of this is the evaluation of total energy $E = \langle\hat{H}\rangle$ via the sampling of local energies $E_L(\mathbf{r}) = \Psi_{\theta}^{-1}(\mathbf{r}) \hat{H}\Psi_{\theta}(\mathbf{r})$.
In the case of $\hat{M}$, total angular momentum in z direction, we write down the analogous expression for its expectation value:
\begin{equation}
\langle \hat{M} \rangle = \frac{\int \Psi_{\theta}^{*}(\mathbf{r})\hat{M} \Psi_{\theta}(\mathbf{r}) d\mathbf{r}^2}{\int \Psi_{\theta}^{*}(\mathbf{r})\Psi_{\theta}(\mathbf{r}) d\mathbf{r}^2} = \frac{\int \lvert \Psi_{\theta}(\mathbf{r}) \lvert^2 \,[\Psi_{\theta}^{-1}(\mathbf{r}) ({-ix\partial_y + iy\partial_x})\Psi_{\theta}(\mathbf{r}) ]\,d\mathbf{r}^2}{\int \lvert \Psi_{\theta}(\mathbf{r}) \lvert^2 d\mathbf{r}^2} = \frac{\int \lvert \Psi_{\theta}(\mathbf{r}) \lvert^2 {M}_L(\mathbf{r})\,d\mathbf{r}^2}{\int \lvert \Psi_{\theta}(\mathbf{r}) \lvert^2 d\mathbf{r}^2} = \langle M_L \rangle
\label{eq:am_exp}\tag{S24}
\end{equation}

Note that angular momentum also involves gradient evaluation, so if one wants to evaluate the angular momentum and the total energy in the same MCMC step, one needs to be careful as KFAC has some donated buffers that effectively prevent us from using the same parameters twice for the evaluation of gradient-related quantities. To circumvent this, the parameters need to be copied, and some trivial operations (such as reshaping back and forth) need to be performed on it so that it survives the donated buffers. No such issues arise for post-training angular momentum measurements.

\begin{figure*}[!h]
    \includegraphics[width=0.98\textwidth]{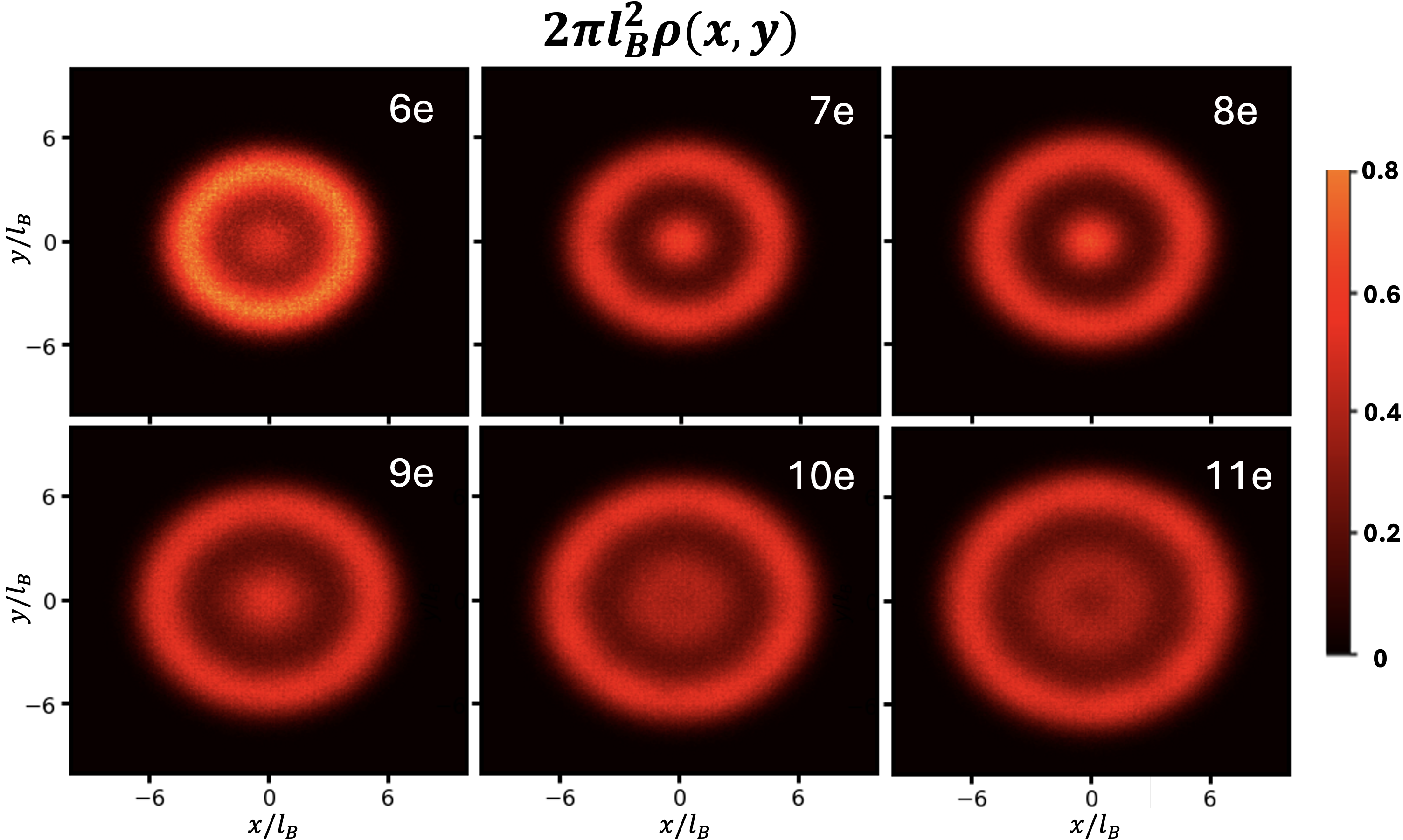}
    \caption{2D charge density for 6-11 electrons with low Landau level mixing ($\lambda = 1/3$). }
    \label{fig:2d_density}
\end{figure*}
%
\section{\large Additional Density Plots 
\label{app:add_data}}

For completeness, we present the 2D charge density for 6-11 electrons in Fig.~\ref{fig:2d_density}. In all cases, Psiformer learns the circular symmetry extremely well. For 6-9 electrons, the charge density is featured by a dot surrounded by a shell, while for more than 9 electrons, the inner dot structure disperses radially outward and forms a inner shell. For 10 and 11 electron cases, we see the charge density profile flattens and becomes more uniform. 

\begin{figure*}[!h]
    \includegraphics[width=0.98\textwidth]{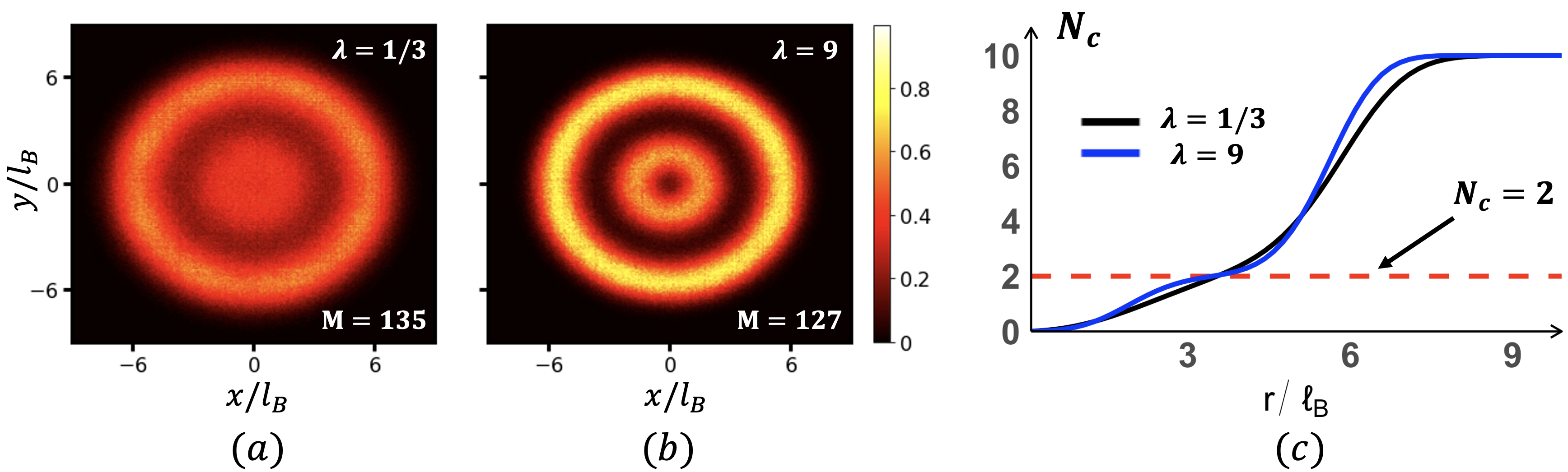}
    \caption{LL mixing induced transition for 10 electrons: (a) shows the density for weak LL mixing $\lambda = 1/3$, while (b) illustrates the strong LL mixing case. (c) gives the cumulative sum of charge in radial direction. }
    \label{fig:10e_LL_mixing}
\end{figure*}

Moreover, we show the transition from FQH liquid to a crystal state for 10 electron case (see Fig.~\ref{fig:10e_LL_mixing}). As shown in Fig.~\ref{fig:10e_LL_mixing}(a) and (b), under strong LL mixing, the electrons are localized to two shells with a decrease of angular momentum from $M = 135$ to $M = 127$, marking a transition away from FQH liquid state. The radial cumulative sum of electrons is plotted in Fig.~\ref{fig:10e_LL_mixing}(c) has a plateau at $N_c = 2$, which indicates that there are two electrons in the inner shell and eight in the outer shell. This transition is discussed in details in the main text.

%
\section{\large Training Hyperparameters \label{app:hyperparams}}

The training hyperparameters is summarized in TABLE~\ref{tab:hyperparams}. Most of the training is done on 4 NVIDIA P100 GPUs or 1 NVIDIA A100 GPU. We notice that TF32, the default precision for A100, sometimes encounters difficulty with training convergence and stability, while single precision is often sufficient for full convergence. Moreover, with single precision in use, P100 tends to take less steps to converge than A100 despite the same task and training hyperparameters. For small systems, a higher initial learning rate $0.02$ together with a smaller delay $20000$ leads to faster convergence. The FNN is built upon FermiNet github repository~\cite{FermiNet_2020} and implemented in JAX~\cite{JAX_2018}. We also noticed that FermiNet has difficulties reaching circular symmetry and quantized angular momentum value, which might be due to its lower expressivity compared to Psiformer.

\begin{table}[!htbp]
\caption{\textbf{Recommended training hyperparameters}}
\setlength{\tabcolsep}{8\tabcolsep}
\begin{tabular}{lSlS} \toprule
   {Hyperparameter} & {Value} & {Hyperparameter} & {Value}  \\ \midrule
   {Network Type} & {Psiformer}  & {Optimizer} & {KFAC}  \\
   {Number of layers} & {2}  & {Number of heads} & {4}  \\
   {Head dimensions} & {64}  & {Layer dimensions} & {256}  \\
   {KFAC norm constraint} & {0.001}  & {KFAC damping} & {0.001} \\
   {Learning rate} & {0.005} & {Batch size} & {6000}  \\ 
   {Delay} & {200000} & {Decay} & {1}  \\
   {Rescale input} & {False} & {Layer norm} &  {False} \\ 
   {Precision} & {FP32} & {MCMC steps between iterations} &  {10} \\ \bottomrule
\label{tab:hyperparams}
\end{tabular}
\end{table}


\section{\large Exact Diagonalization Calculation for Disk Geometry}
We perform lowest Landau level exact diagonalization in the symmetric gauge, where the single-particle orbitals are labeled by their angular momentum quantum number $l$. We do not truncate the single-particle basis by hand. Instead, the allowed LLL orbital configurations are constrained by the total angular momentum, which is a conserved quantity. Our ED calculation assumes the ground state has $M=M_L$, which is independently confirmed by our FNN results.  
For additional details on ED calculations, please see the Supplemental Material of \cite{2407.09204}.
%
%
%

\end{document}